\documentclass[11pt,a4paper]{article}

\let\ssection=\section
\renewcommand{\section}{\setcounter{equation}{0}\ssection}

\title{Relativistic Chaplygin gas
 with field-dependent Poincar\'e symmetry
\\
}
\author{
M.~Hassa\"{\i}ne and P.~A.~Horv\'athy\\
Laboratoire de Math\'ematiques et de Physique Th\'eorique\\
Universit\'e de Tours, Parc de Grandmont\\
F-37200 TOURS (France)\\
}

\def\IR{{\bf R}}
\def\smallover#1/#2{\hbox{$\textstyle{#1\over#2}$}} %
\def\O{{\rm O}}
\def\parag{\hfil\break} 
\def\kikezd{\parag\underbar}
\def\const{{\rm const}}
\def\boxit#1{
\vbox{\hrule\hbox{\vrule\kern4pt
\vbox{\kern5pt#1\kern5pt}\kern4pt\vrule}\hrule}
} 
\newcommand\vnabla{\mathop{\vec\nabla}\nolimits}
\newcommand\p{\mathop{\partial}\nolimits}

\begin{document}
\maketitle
\begin{abstract}
The relativistic generalization of the Chaplygin gas,
put forward by Jackiw and Polychronakos,
is derived in Duval's
Kalu\-za-Klein framework, using a universal quadratic
Lagrangian.
Our framework yields a simplified proof
of the field-dependent Poincar\'e symmetry
Our action is
related to the usual Nambu-Goto action [world volume]
of $d$-branes in the same way as the Polyakov and the
Nambu action are in strings theory.
\end{abstract}

\vskip2mm

\noindent
{\sl Lett. Math. Phys}. {\bf 57}, pp. 33-40 (2001).

\section{Introduction}

In the light-cone gauge, a relativistic $d$-brane moving in $(d+1,1)$
dimensi\-onal Minkowski space yields a $(d,1)$ dimensional
 isentropic and irrotational fluid,
called the  Chaplygin gas \cite{BHI, JACKPO}. This fluid obeys
the equations of motion
 \begin{equation}
\partial_{t} R+\vnabla\cdot\big( R\vnabla\Theta\big)=0,
\qquad
\partial_t\Theta+\frac{1}{2}\big(\vnabla\Theta\big)^2=-\frac{dV}{dR}.
\label{eqmotion}
\end{equation}
where $ R(x,t)\geq0$ is the density, $\Theta(x,t)$ is the
velocity potential, and
\begin{equation}
V(R)=\frac{c}{R},
\qquad
c=\const.
\label{membpot}
\end{equation}
More generally, one can consider the polytropic potential
$
V=cR^{n},
$
where ${n}$ is a real constants.
In this Letter we shall mostly restrict ourselves
to the membrane case ${n}=-1$.

One of the surprising features of the Chaplygin system is its
large, non-linearly realized, field-dependent symmetry~:
its manifest $(d,1)$-dimensional galilean symmetry extends in fact
into a curious
$(d+1,1)$-dimensional Poin\-car\'e dynamical symmetry
 \cite{BHI, JACKPO, JEV, BJ}.

 In \cite{HH},  this field-dependent
Poincar\'e symmetry was linearized by unfolding the system into a
``Kaluza-Klein'' spacetime $M$,
obtained by adding a coordinate $s$ to non
relativistic space and time, $x$ and $t$.  Then
  all the field-dependent symmetries
became Poincar\'e transformations of
$(d+1,1)$ dimensional
Minkowski space $M$ with metric $dx^2+2dtds$.
($t$ and $s$ are hence light-cone coordinates).

So much for the kinematics. Interestingly,
 the non-relativistic Chaplygin model could itself
 be derived
by {\it lightlike} projection from
$(d+1,1)$ dimensional Minkowski space $M$,
 \cite{HH}\footnote{According
to our conventions, $i, j$ are spatial
indices, $\alpha,\beta,\dots$ refer
to coordinates on ordinary spacetime, and $\mu,\nu,\dots$ refer
to the extended ``Kaluza-Klein'' spacetime, $M$.}.
Let us indeed consider two
real fields $\varrho$ and $\theta$ and the potential
$V(\varrho)=\lambda/\varrho$, and posit the equations
(\ref{KGeq}) below.

 Then, suitably defining the projected fields (see \cite{HH} for
details) we get
  (\ref{eqmotion}) with the potential (\ref{membpot}).
The {\it manifest}
$(d+1,1)$-dimensional
Poincar\'e symmetry of the higher-dimensional model
can be shown, furthermore, to be preserved by the conditions
(\ref{thecondition}), proving
 the dynamical Poincar\'e symmetry.
  Previous proofs use either a tedious
direct calculations
\cite{JEV, BJ} or follow by a rather tricky reduction from
membrane theory \cite{BHI, BHII, JACKPO}.

\goodbreak
 Recently \cite{JACKPO},
Jackiw and Polychronakos presented a relativistic generalization
of the Chaplygin gas, with Lagrange density
\begin{equation}
L^{\hbox{JP}}=
\Theta\,\partial_\tau R-
\sqrt{R^2c^2+a^2}\sqrt{c^2+(\vec\nabla\Theta)^2},
\label{JPLag}
\end{equation}
where $\tau$ denotes the relativistic time,
$\Theta$ is the momentum  potential,
$\rho$ the density,
 and the constant $a$ is the interaction strength.
This specific form is chosen so that
 the non-relativistic Chaplygin model
is recovered in the limit $c\to\infty$.
In what follows we set $c=1$
and focus our attention to
the relativistic model.
The equations of motion associated to (\ref{JPLag}) read
\begin{equation}
    \left\{\begin{array}{c}
	\partial_{\tau}R+\vec\nabla\cdot
	\left(\vec\nabla\Theta\,\sqrt{\displaystyle\frac{R^2+a^2}
	{1+(\vec\nabla\Theta)^2}}\right)=0,\hfill\\[4mm]
	\partial_{\tau}\Theta+R\sqrt{
	\displaystyle\frac{1+(\vec\nabla\Theta)^2}
	{R^2+a^2}}=0.\hfill\\
     \end{array}\right.
     \label{JPeqmot}
\end{equation}

\goodbreak
The manifest $(d,1)$-dimensional Poincar\'e symmetry of
 (\ref{JPLag}) extends, just like for its
non-relativistic counterpart, to a field-dependent
$(d+1,1)$-dimensio\-nal Poincar\'e dynamical symmetry.
The additional symmetries are
time repara\-metri\-za\-tion, $\widetilde{x}=x$,
\begin{equation}
\begin{array}{c}
\widetilde{\tau}=
\displaystyle\frac{\tau}{\cosh\omega}+\Theta(\widetilde{\tau},x)
\tanh\omega,\hfill\\[3mm]
\widetilde{\Theta}=\displaystyle\frac{\Theta(\widetilde{\tau}, x)}{\cosh\omega}
-\tau\tanh\omega,\hfill
\end{array}
\label{Rtime}
\end{equation}
and space reparametrization, $\widetilde{\tau}=\tau$,
\begin{equation}
\begin{array}{c}
\widetilde{x}=x
-
\displaystyle\hat{\gamma}\,
{\Theta}(\tau,\widetilde{x})\tanh\gamma
+
\displaystyle\hat{\gamma}
\big(\displaystyle\hat{\gamma}\cdot x\big)
\left(\displaystyle\frac{1-\cos\gamma}{\cos\gamma}\right),\hfill\\
[2.6mm]
\tilde{\Theta}=\displaystyle\frac{\Theta(\tau,\widetilde{x})-
(\hat{\gamma}\cdot x)\sin\gamma}
{\cosh\gamma},\hfill
\end{array}
\label{Rspace}
\end{equation}
where $\gamma=\vert\vec{\gamma}\vert$ and
$\hat{\gamma}=\vec{\gamma}/\gamma$.

The aim of this Note is to derive also
the relativistic model of Jackiw and Polychronakos from the {\it same
universal model} as the non-relativistic Chaplygin system, but using
{\it spacelike} rather than lightlike projection.
First, we provide a similar interpretation
of space and time reparametri\-zations
as isometries of an extended space.
Next we consider a
non-linear Klein-Gordon system and point out that
its ``Madelung'' transcription \cite{MADELUNG} yields, for a suitable choice
of the potential, the universal model
(\ref{KGeq}) and (\ref{Bargaction}) respectively, refered to above.
This will also demonstrate the field-dependent Poincar\'e dynamical
symmetry of both Chaplygin systems.

The relation to branes is discussed in Section 4.
\goodbreak

\section{Unfolding}

Let us start with the time reparametrizations, (\ref{Rtime}).
Following the same recipe as in the non-relativistic case,
let us add the new coordinate
$
\sigma=-\widetilde{\Theta}
$
$\Longrightarrow$
$\widetilde{\sigma}=-\Theta.
$
 Then (\ref{Rtime}) yields $\widetilde{x}=x$,
\begin{equation}
\begin{array}{c}
\widetilde{\tau}=\cosh\omega\,\tau-\sinh\omega\,\sigma,\hfill
\\
\widetilde{\sigma}=\cosh\omega\,\sigma-\sinh\omega\,\tau.\hfill
\end{array}
\label{LRtime}
\end{equation}
which is in fact a Lorentz transformation in the $\sigma$ direction
of Minkowski space with metric
$-d\tau^{2}+dx^{2}+d\sigma^{2}$.
($\tau$ is hence timelike and $\sigma$ spacelike).
Switching to the light-cone coordinates
$
t=\frac{-\tau+\sigma}{2},
\,
s=\frac{\tau+\sigma}{2},
$
(\ref{LRtime}) becomes furthermore
the non-relativistic time dilation
$\widetilde{x}=x$, $\widetilde{t}=e^\delta t$,
$\widetilde{s}=e^{-\delta}s$ \cite{HH}.

Space reparametrizations admit a similar interpretation.
Applying again our rules,  (\ref{Rspace})
unfolds as a rotation $d+1$-dimensional space, $\tilde{\tau}=\tau$,
\begin{equation}
\begin{array}{c}
\widetilde{x}=x
-
\displaystyle\hat{\gamma}\sin\gamma\,\sigma
-
\displaystyle\hat{\gamma}
\big(\displaystyle\hat{\gamma}\cdot x\big)
(1-\cos\gamma),\hfill\\[1mm]
\tilde{\sigma}=\cos\gamma\,\sigma-
(\displaystyle\hat{\gamma}\cdot x\big)\sin\gamma.\hfill
\end{array}
\label{LRspace}
\end{equation}

Interestingly, a $(d,1)$ dimensional Lorentz boost
 lifted to our extended space,
$\tilde{\sigma}=\sigma$,
\begin{equation}
\begin{array}{c}
\widetilde{x}=x
+
\displaystyle\hat{\beta}\sinh\beta\,\tau
-
\displaystyle\hat{\beta}
\big(\displaystyle\hat{\beta}\cdot x\big)
(1-\cosh\beta),\hfill\\[1mm]
\tilde{\tau}=\cosh\beta\,\tau+\big(\displaystyle
\hat{\beta}\cdot x\big)\sinh\beta,\hfill\\[1mm]
\end{array}
\label{LLorentz}
\end{equation}
($\beta=\vert\vec\beta\vert,\, \hat{\beta}=\vec{\beta}/\beta$)
is related to the space reparametrization by the interchange
of $\tau$ and $\sigma$ and by changing $\gamma$ into $i\beta$.
(In the non-relativistic case,
 ``antiboosts'' and
 galilean boosts are related
interchanging the light-cone coordinates $s$ and $t$ \cite{HH}).

\section{Dynamics}

Let us consider a Klein-Gordon field $\psi$
on $(d+1,1)$-dimensional Minkowski space,
\begin{equation}
\p_{\mu}\p^{\mu}\psi
=2\frac{d\tilde{V}}{d\psi^*}
\label{KleinGordon}
\end{equation}
where $\tilde{V}=\tilde{V}(\vert\psi\vert^{2})$ is some potential.
Now, in analogy with the  well-known hydrodynamical transcription of
non-relativistic quantum mechanics due to Madelung \cite{MADELUNG},
we write
$\psi=\sqrt{\varrho}\,e^{i\theta}$ to get
\begin{equation}
\left\{\begin{array}{c}
\partial_{\mu}\big(\varrho\,\partial^{\mu}\theta\big)=0,\hfill\\
[3mm]
\frac{1}{2}\partial_{\mu}\theta\partial^{\mu}\theta
=-\displaystyle\frac{\delta V}{\delta\varrho}
\end{array}\right.
\label{KGeq}
\end{equation}
where $\delta V/\delta\varrho$ is the variational derivative, and
\begin{equation}
V=\tilde{V}+\frac{1}{8}\frac{\p_{\mu}\varrho\p^\mu\varrho}{\varrho}
\label{effpot}
\end{equation}
is an effective potential involving the original one, $\tilde{V}$,
 plus a ``quantum'' contribution. If we chose,
following Bazeia and Jackiw \cite{BJ}, $\tilde{V}$ so that it cancels
the second term,
$\tilde{V}=V-\p_{\mu}\varrho\p^\mu\varrho/8\varrho$, (\ref{KGeq})
reduces precisely to (\ref{KGeq}).
Similarly, the  action
from which the non-linear Klein-Gordon equation (\ref{KleinGordon}) is
derived,
\begin{equation}
\int d^{d+2}x\left\{-\frac{1}{2}\p_{\mu}\psi\p^{\mu}\psi^*
-\tilde{V}\right\}
\end{equation}
is converted under the Madelung transcription  into
\begin{equation}
\int d^{d+2}x\left\{
-\frac{1}{2}\varrho\,\partial_\mu\theta\,\partial^\mu\theta-V
\right\}
\label{Bargaction}
\end{equation}
with $V$ the effective potential (\ref{effpot}).
The Euler-Lagrange equations are
precisely (\ref{KGeq}).
This provides a physical interpretation of the ``lifted system''
(\ref{KGeq})--(\ref{Bargaction}).

A word of caution, however: The upper equation in (\ref{KGeq})
can  be viewed as a continuity equation {\it not} for $\varrho$ but for
$
(i/2)(\psi^*\p_{\tau}\psi-\psi\p_{\tau}\psi^*)
=
\varrho\p^{\tau}\theta
$
which, as it is well-known
\cite{BJORKENDRELL}, not a positive definite expression.
Thus, it can not be viewed as ``particle density'', only,
in the best case, a ``charge density''. (This same
interpretational problem concerns the
relativistic system of Jackiw and Polychronakos (\ref{JPLag}) and
(\ref{JPeqmot}).)

We now derive
the relativistic model using instead {\it spacelike} projection.
Let us hence consider the relativistic
coordinates $x, \tau, \sigma$
on Minkowski space. Then, generalizing the rules in \cite{HH}
to the relativistic context,
we define the fields
$\Theta$ and $\rho$ by the conditions
\begin{equation}
\begin{array}{c}
\theta\big(x,\tau,-\Theta(x,\tau)\big)=0,\hfill\\[1.6mm]
\rho(x,\tau)=\varrho\big(x,\tau,-\Theta(x,\tau)\big)
\p_{\sigma}\theta(x,\tau,-\Theta(x,\tau).
\end{array}
\label{thecondition}
\end{equation}
\goodbreak

 The first here is an implicite equation for $\Theta$, viewed as a
 field on ordinary space-time; once this latter has been determined,
 it can be used to define the projected field
 $\rho$. It is important to observe that this procedure is
 only consistent with the potential
 $1/\varrho$ \cite{HH}.
Then the ``universal''
 equations of motion (\ref{KGeq}) project,
  for $V\propto 1/\varrho$ only \cite{HH},
to the manifestly $(d,1)$-dimensional Poin\-car\'e invariant
expressions
\begin{equation}
    \left\{\begin{array}{c}
    \partial_{\tau}(\rho\partial^{\tau}\Theta)
    +\vec\nabla\cdot\big(\rho\vec\nabla\Theta\big)=0,\hfill
    \\[2mm]
    -(\partial_{\tau}\Theta)^2+(\vec\nabla\Theta)^2+1=
    \displaystyle\frac{2\lambda}{\rho^2}.\hfill\\
\end{array}\right.
\label{projeqmot}
\end{equation}

What is the physical interpretation of these  equations ?
In analogy with the non-relativistic case, we would like to
interpret the upper equation here as a continuity equation,
i. e., a conservation equation $\p_{\alpha}J^\alpha=0$
for the non-relativistic current
\begin{equation}
J^{\tau}=\rho\p^{\tau}\Theta,
\qquad
\vec{J}=\frac{\vnabla\Theta}{\p^{\tau}\Theta}.
\label{conscurrent}
\end{equation}

Now the point is that, trading $\rho$ for
\begin{equation}
    \begin{array}{c}
    R\equiv J^{\tau}=
    \varrho\big(x,\tau,-\Theta(x,\tau)\big)\p^{\tau}\theta
    \big(x,\tau,-\Theta(x,\tau)\big)
\\[2mm]
=
\varrho\big(x,\tau,-\Theta(x,\tau)\big)\p_{\sigma}\theta
\big(x,\tau,-\Theta(x,\tau)\big)\p^{\tau}\Theta,
\end{array}
 \label{Rdef}
\end{equation}
 (\ref{projeqmot}) become precisely
the equations of Jackiw and Polychronakos in (\ref{JPeqmot})
with $a=\sqrt{2\lambda}$. In the same spirit, the
universal action (\ref{Bargaction}) becomes, under
(\ref{thecondition}),
\begin{equation}
L=\frac{1}{2}\rho
\left((\partial_{\tau}\Theta)^2-(\vec\nabla\Theta)^2-1\right)
-\frac{\lambda}{\rho}.
\label{projaction}
\end{equation}
Eliminating $\rho$ in favor of $R$ this nice quadratic expression
becomes, furthermore, the square-root action (\ref{JPLag}).

In conclusion, we have derived the relativistic system of Jackiw and
Polychronakos,
\cite{JACKPO}
by spacelike projection from our
universal model (\ref{KGeq}).
(Remember that in the non-relativistic case one had to use lightlike
projection).
Beyond its esthetical value, our construction has the advantage that
the  $(d+1,1)$ dimensional dynamical Poincar\'e symmetry
becomes a simple consequence of the manifest geometric
Poincar\'e invariance of the universal model.
This can be shown along the same lines as in \cite{HH}.

A further advantage is that the conserved quantity
associated to an infinitesimal Poincar\'e
transformation $(X^\mu)$ of $M$ is readily found using the [symmetric]
energy-momentum tensor of (\ref{Bargaction}) constructed in \cite{HH},
\begin{equation}
\begin{array}{c}
 Q=\displaystyle{\int\frac{T^{\tau}_{\ \mu }X^\mu}{\partial_{\sigma}\theta}\,
 d^dx},\hfill
 \\[3.4mm]
T_{\mu\nu}=-\varrho\,\partial_\mu\theta\partial_\nu\theta
+g_{\mu\nu}\left(
\smallover1/2\rho\partial_\omega\theta\partial^\omega\theta
+V(\varrho)\right).\hfill
\end{array}
\label{emom}
\end{equation}
These formulae yield
\begin{equation}
\begin{array}{cc}
{\cal H}=\smallover1/2\rho\big[
(\partial_\tau\Theta)^2+(\vec\nabla\Theta)^2+1\big]
+\displaystyle\frac{\lambda}{\rho},\quad
\hfill
&\hbox{energy}\hfill
\\[2mm]
{\cal P}_i=-\rho\partial_{i}\Theta\,\partial_{\tau}\Theta,\hfill
&\hbox{momentum}\hfill
\\[2mm]
{\cal N}=-\rho\partial_{\tau}\Theta,\hfill
&\hbox{relat. ``number''}\hfill
\\[2mm]
{\cal D}={\cal H}\Theta+{\cal N}\tau,\hfill
&\hbox{time reparametrization}
\\[2mm]
{\cal G}_i=x_i{\cal N}+\Theta{\cal P}_i
\hfill&\hbox{space reparametrization}\hfill
\end{array}
\label{ourconserved}
\end{equation}
which become, inserting $R$,
exactly the conserved quantities in \cite{JACKPO}.

Somewhat paradoxically, both {\it relativistic} systems,
(\ref{JPLag}) and (\ref{projaction}), are also Galilei-invariant,
simply because the $(d,1)$ dimensional Galilei group is a subgroup
of the Poincar\'e group in $(d+1,1)$ dimensions.
Applying our rules backwards,
for a galilean boost
we get, e. g., the field-dependent action
\begin{equation}
\begin{array}{c}
\widetilde{x}=x-\smallover1/2{\alpha}\tau
-\smallover1/2{\alpha}\widetilde{\Theta},\hfill
\\[1.5mm]
\widetilde{\tau}=\big(1+\smallover1/4{\alpha}^2\big)\tau
-{\alpha}\cdot x
+\smallover1/4{\alpha}ý^2\widetilde{\Theta},\hfill
\\[1.5mm]
\Theta=\widetilde{\Theta}\big(1+\smallover1/4{\alpha}^2\big)
+{\alpha}\cdot x-\smallover1/4{\alpha}^2\tau
+\widetilde{\Theta}\big(1+\smallover1/4{\alpha}^2\big).\hfill
\end{array}
\end{equation}

\section{Relation to d-branes}

Our framework here is closely related to the so-called
 non-parametric representation of $d$ branes \cite{BHII}.
Our ``vertical'' variable $\sigma$
({\it alias} the field $-\Theta$)
is in fact the $z$ coordinate
of the $d$-brane
propagating in $(d+1,1)$ dimensional Minkowski space,
and our ``lifted'' field $\theta$ is (minus) their $u$,
the function
whose level sets describe the $d$-brane as
$\theta(x,\tau,z(x,\tau))=0$,
--- which is our first condition in (\ref{thecondition}).

In terms of $\theta$, the motion of the $d$-brane is governed by the
action
\begin{equation}
\int\sqrt{\partial_{\mu}\theta\,\partial^{\mu}\theta}\,d^{d+2}x,
\label{BHaction}
\end{equation}
whose equations of motion read
\begin{equation}
    \partial_{\mu}\left(\frac{\partial^\mu\theta}
    {\sqrt{\partial_{\nu}\theta\,\partial^{\nu}\theta}}\right)=0.
    \label{BHeqmot}
\end{equation}
The integrand here is in fact the ``Nambu''
world volume of the $d$-brane \cite{Hoppe},
\begin{equation}
\sqrt{\partial_{\mu}\theta\,\partial^{\mu}\theta}=
\sqrt{\det(G_{\alpha\beta})},\qquad
G_{\alpha\beta}=\partial_\alpha X^\mu\partial_\beta X_\mu.
\label{wvol}
\end{equation}

The point is that one can get rid of the square root,
just like for a free relativistic particle.
(This latter can
 be described either by the usual invariant length action
 $-m\displaystyle\int\!\sqrt{-\dot{x}^{2}}\,d\tau$,
or by a quadratic action plus a
constraint, when  an auxiliary variable is added \cite{GSW}).
Let us hence enlarge our pure scalar theory involving $\theta$ alone by
introducing an auxiliary field we call $\rho$.
Then (\ref{BHeqmot}) is readily seen to imply
the first equation in (\ref{KGeq}); but both
equations (\ref{KGeq}) derive from our quadratic
Lagrangian (\ref{Bargaction}).
Conversely, inserting $\rho$ into our action and equations of
motion, (\ref{BHaction}) and (\ref{BHeqmot}) are recovered.
(The two-dimensional  analog is
string theory, where the quadratic Polyakov action can be
used instead of the Nambu-Goto expression \cite{POLYAKOV, GSW}).

\section{Discussion}

While the Lagrangian (\ref{JPLag})
is first-order in the time derivative
and the Hamiltonian
$\sqrt{R^2+a^2}\,
\sqrt{\left(\vec\nabla\Theta\right)^2+1}$
contains ugly square roots,
our expressions are quadratic, as in ordinary
relativistic scalar field theory.
The two expressions are equivalent;
our quadratic expression could have some advantage when the
quantization of the system is considered.
Let us
emphasise that this approach not only considerably
simplifies the proof of the dynamical Poincar\'e symmetry, but also
explains its  rather mysterious origin.
The possibility of having differently-looking but
still equivalent
systems corresponds to the freedom of chosing
the kinetic term \cite{JACKPO}.

The strange fact  recognized by Jackiw and Polychronakos
is that  the  the non-relati\-vis\-tic Chaplygin gas is simultaneously
the $c\to\infty$ limit, and  also equivalent
to their relativistic model \cite{JACKPO}.
This can also be seen in our framework~: deforming the space-like
fibration into lightlike amounts, on the one hand,  to taking the
non-relativistic limit \cite{DHP}. On the other hand,
(\ref{thecondition}) is merely the definition of the projected
fields and does not impose any restriction. The two
systems are hence equivalent through the universal model.


Let us mention, in conclusion, that
our formalism can also be used to
study the conformal properties of
 gas dynamics \cite{poly}. For the adiabatic potential
$V(\varrho)\propto\varrho^n$, the action (\ref{Bargaction}) is readily
seen to be invariant w. r. t. the $(d+1,1)$ dimensional conformal
group $\O(d+1,2)$
if and only if the polytropic exponent is
\begin{equation}
    n=1+\frac{2}{d}.
\end{equation}
(This can also be seen from the trace condition $T^\mu_{\ \mu}=0$
of the energy-momentum tensor (\ref{emom})).

In the free case, $\O(d+1,2)$ is a [field-dependent]
symmetry also for the reduced system
\cite{HH}. For $V\neq0$, however,
the potential is only consistent with equivariance,
\begin{equation}
    \begin{array}{ccc}
    \partial_{\sigma}\varrho=0\hfill
    &\Longrightarrow
    &\varrho=\rho(x,\tau),\hfill\\[1.4mm]
    \partial_{\sigma}\theta=1\hfill
    &\Longrightarrow\hfill
    &\theta=\Theta(x,\tau)+\sigma,\hfill
\end{array}
\label{equivariance}
\end{equation}
rather than with the generalized condition (\ref{thecondition}).
Equivariance reduces, however, the $(d+1,1)$ dimensional conformal
symmetry to its mere
\big[(d,1)-Poincar\'e\big]$\times\IR$ subgroup, the $\IR$ representing the
vertical
translations, whose associated conserved quantity is the ``number''
$N$.
Let us recall that in the non-relativistic case the corresponding
subgroup is the $(d,1)$ dimensional Schr\"odinger group
\cite{JNH, HH, poly}.

\goodbreak
\kikezd{Acknowledgements}.
We are indebted to Professors C. Duval, G. Gibbons,  R. Jackiw
and J.-W. van Holten
 for their interest, and to J. Hoppe for discussions.

\goodbreak

\end{document}